\documentclass[journal]{IEEEtran} \newcommand{\figsza}{0.99} \newcommand{\figszb}{0.99} \newcommand{\figszc}{0.92}\newcommand{\figszd}{0.92}\newcommand{\figsze}{0.99}

\usepackage[cmex10]{amsmath}
\usepackage{amsfonts}
\usepackage{amssymb}
\usepackage{amsopn}
\usepackage{graphicx}
\usepackage{subfigure}
\usepackage{epsfig}
\usepackage{color}
\usepackage{setspace}
\usepackage{bbm}

\newtheorem{theorem:ideal:QCQP}{Theorem}
\newtheorem{theorem:conn}[theorem:ideal:QCQP]{Proposition}
\newtheorem{theorem:lowerbd}[theorem:ideal:QCQP]{Proposition}
\newtheorem{theorem:all:link}[theorem:ideal:QCQP]{Proposition}
\newtheorem{theorem:cycle}[theorem:ideal:QCQP]{Proposition}
\newtheorem{theorem:inequality}[theorem:ideal:QCQP]{Lemma}

\graphicspath{{fig/}}

\newcommand{\tr}{\text{Tr }}
\newcommand{\bo}[1]{\boldsymbol{#1} }

\newcommand{\indicator}[1]{\mathbbm{1}_{\left[ {#1} \right] }}
\makeatletter
\def\blfootnote{\xdef\@thefnmark{}\@footnotetext}
\makeatother

\begin{document}

%
\title{\huge{On Linear Coherent Estimation with Spatial Collaboration}}

\author{Swarnendu~Kar,~\IEEEmembership{Student~Member,~IEEE,}
        and~Pramod~K.~Varshney,~\IEEEmembership{Fellow,~IEEE}
\thanks{S. Kar and P. K. Varshney are with the Department
of Electrical Engineering and Computer Science, Syracuse University, Syracuse,
NY, 13244 USA. E-mail: \{swkar,varshney\}@syr.edu.}
\thanks{This is a manuscript under preparation and will be submitted to IEEE Transactions on Signal processing. Part of this work has been accepted for publication in Proceedings of ISIT-2012, IEEE International Symposium of Information Theory, July 1--6, 2012, Cambridge, MA, USA. }
\thanks{This research was partially supported by the National Science Foundation under Grant No. $0925854$ and the Air Force Office of Scientific Research under Grant No. FA-9550-10-C-0179. }
}

\maketitle

\begin{abstract}
We consider a power-constrained sensor network, consisting of multiple sensor nodes and a fusion center (FC), that is deployed for the purpose of estimating a common random parameter of interest. In contrast to the distributed framework, the sensor nodes are allowed to update their individual observations by (linearly) combining observations from neighboring nodes. The updated observations are communicated to the FC using an analog amplify-and-forward modulation scheme and through a coherent multiple access channel. The optimal collaborative strategy is obtained by minimizing the cumulative transmission power subject to a maximum distortion constraint. For the distributed scenario (i.e., with no observation sharing), the solution reduces to the power-allocation problem considered by \emph{Xiao et. al.} \cite{Xiao08}. Collaboration among neighbors significantly improves power efficiency of the network in the low local-SNR regime, as demonstrated through an insightful example and numerical simulations.
\end{abstract}

\section{Introduction}
\begin{figure}[htb]
\vspace{-0.1in}
\begin{center}
    \includegraphics[width=\figsza \columnwidth]{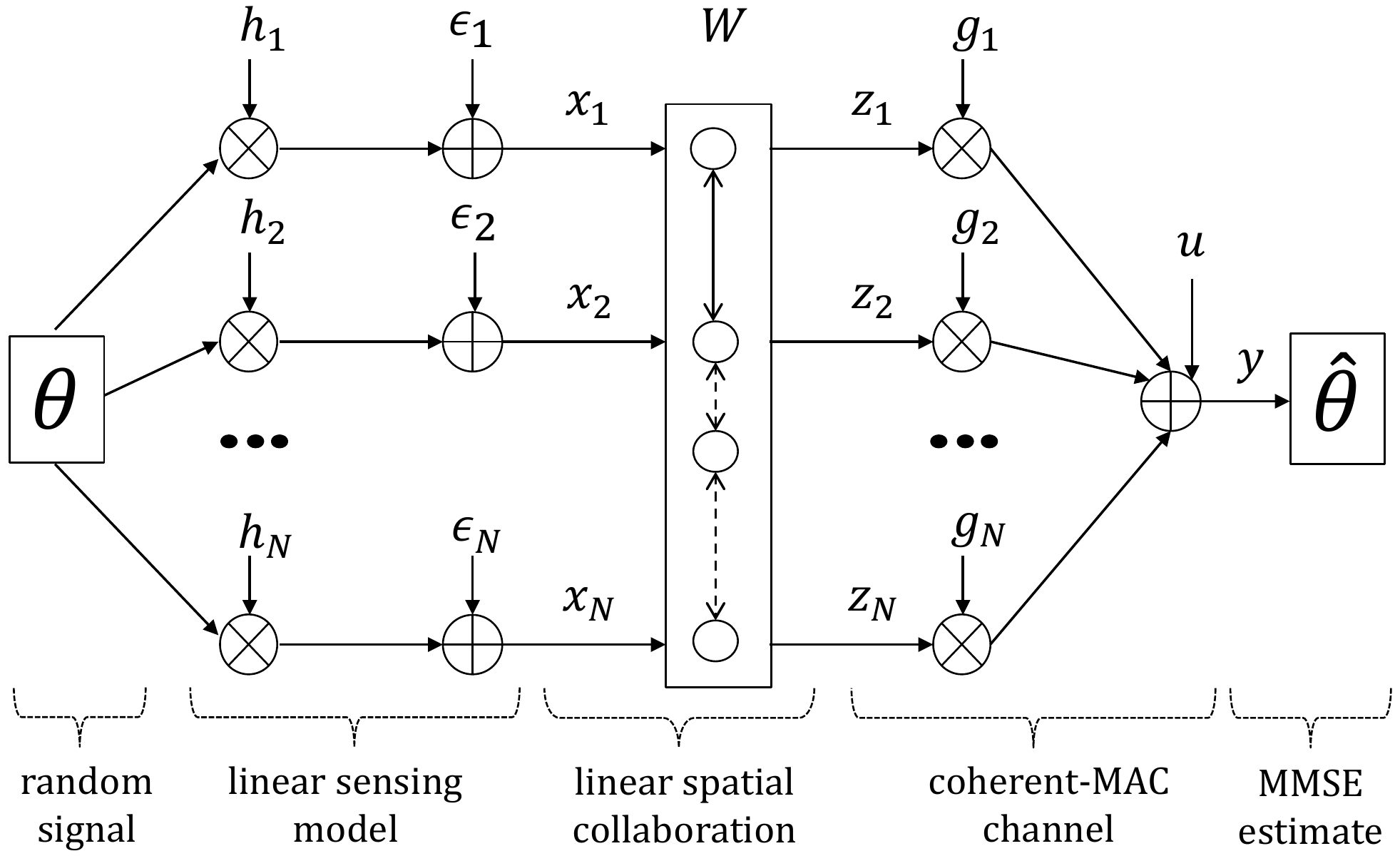}
\vspace{-0.24in}
  \caption{Sensor network performing collaborative estimation.}
  \label{fig:schematic}
  \end{center}
\vspace{-0.2in}
\end{figure}

Wireless sensor networks consist of spatially distributed battery-powered sensors that monitor certain environmental conditions and often cooperate to perform specific signal processing tasks like detection, estimation and classification \cite{Akyildiz02}. In this paper, we consider a network that is deployed for the purpose of estimating a common random parameter of interest. After observing noisy versions of the parameter, the sensors can share their observations among other neighboring nodes, an act referred to as \emph{collaboration} in this paper (following \cite{Fang09}). The observations from all the neighbors are linearly combined and then transmitted to the fusion center (FC) through a coherent MAC channel. The FC receives the noise-corrupted signal and makes the final inference. The schematic diagram of such a system is shown in Figure \ref{fig:schematic} (we will introduce the notations and describe each block later in Section \ref{sec:probform}).

The individual sensor nodes are battery powered and hence the network, as a whole, is highly power limited. In the absence of a power limit, the sensors could ideally collaborate with all the other nodes, make the inference in the network, and transmit the estimated parameter to the FC without any further distortion (by using infinite transmission power). However, in the presence of a strict power constraint, both collaboration and transmission have to be performed judiciously, so as to maximize the quality of inference at the FC. In this paper, we study the tradeoff between cumulative transmission power and the quality of inference for a given collaborative neighborhood. We assume \emph{cost-free collaboration}, i.e., the power required to share observations within the neighborhood is negligibly small compared to the power required to communicate with the FC. In an extended version of this paper, we would address the more general problem where collaboration incurs a finite cost.

In the absence of collaboration, this problem is the same as \emph{distributed estimation}, which has been extensively researched - both from analog \cite{Xiao08},\cite{Cui07} and digital \cite{Rib06},\cite{Li09} encoding perspectives. When the parameter to be estimated is a scalar, as in our case, much of the problem formulation is similar to \emph{distributed beamforming in relay networks} \cite{Havary08},\cite{Jing09}. However, research regarding collaborative estimation is relatively nascent. When the transmission channels are orthogonal and cost-free collaboration is possible within a fully connected sensor network, it has been shown in \cite{Fang09} that it is optimal to perform the inference in the network and use the best available channel to transmit the estimated parameter. \emph{In this paper, we study the optimal collaboration design for the partially connected network and coherent MAC channel}.

\section{Problem Formulation} \label{sec:probform}
We consider the scenario where the parameter of interest is a scalar random variable with known statistics, specifically, Gaussian distributed with zero mean and variance $\eta^2$. The observations at the sensor nodes $n=1,2,\ldots,N$ are governed by the linear model $x_n = h_n \theta+\epsilon_n$, where $h_n$ is the source attenuation and $w_n$ is the measurement noise. Let $\bo h=[h_1,h_2,\ldots,h_N]^T$. The measurement noise vector $\bo \epsilon=[\epsilon_1,\epsilon_2,\ldots,\epsilon_N]^T$ is assumed to be zero-mean, Gaussian with (spatial) covariance $\mathbb E[\bo \epsilon\bo \epsilon^T]=\bo \Sigma$. \emph{Perfect knowledge of the observation model parameters $\{h_n\}_{n=1}^N$ and $\bo \Sigma$ is assumed.} In vector notation, the observation model is
\begin{align}
\bo x = \bo h \theta + \bo \epsilon, \quad \theta\sim\mathcal N(0,\eta^2), \bo \epsilon\sim\mathcal N(0,\bo \Sigma), \label{def:x}
\end{align}
where $\bo x=[x_1,x_2,\ldots,x_N]^T$ denotes the vector of observations.

We consider an extension of the analog amplify-and-forward scheme as our encoding and modulation framework for communication to the fusion center. In the basic amplify-and-forward scheme, each node transmits a weighted version of its own observation, say $W_n x_n$, with resulting power $W_n^2 \mathbb E[x_n^2]$. Such a scheme is appealing and often-used (e.g., \cite{Cui07}, \cite{Xiao08}, \cite{Fang09}) due to two reasons, 1) \emph{Uncoded nature:} Does not require block coding across time and hence efficient for low-latency systems, 2) \emph{Optimal:} For a memoryless Gaussian source transmitted through an AWGN channel (Figure \ref{fig:schematic} with $N=1$), an amplify-and-forward scheme helps achieve the optimal power-distortion tradeoff in an information-theoretic sense (see Example $2.2$ in \cite{Gastpar02}). The optimality of linear coding has also been established \cite{Gastpar03} for distributed estimation over a coherent MAC (Figure \ref{fig:schematic} without spatial collaboration) when the observation noise is spatially uncorrelated.

Let the availability of collaborative links among the various nodes be represented by the $N\times N$ adjacency matrix (not necessarily symmetric) $\bo A$, where $A_{ij}\in \{0,1 \}$. An entry $A_{ij}=1$ signifies that node $j$ shares its observations with node $i$. Sharing of this observation is assumed to be realized through a reliable communication link that  consumes power $C_{i,j}$, regardless of the actual value of observation. The $N\times N$ matrix $\bo C$ describes all the costs of collaboration among various sensors and is assumed to be known. Since each node is trivially connected to itself, $A_{ii}=1$ and $C_{ii}=0$. We denote the set of all $\bo A$-sparse matrices as
\begin{align}
\mathcal S_A\triangleq \{\bo W\in\mathbb R^{N\times N}:W_{ij}=0 \text{ if } A_{ij}=0 \}.
\end{align}
Corresponding to an adjacency matrix $\bo A$ and an $\bo A$-sparse matrix $\bo W$, we define \emph{collaboration} in the network as individual nodes being able to linearly combine local observations from other collaborating nodes, $z_n=\sum_{j:A_{nj}=1} W_{nj} x_j$. In effect, the network is able to achieve a one-shot spatial transformation $\bo W: \bo x\rightarrow \bo z$ of the form
\begin{align}
\bo z=\bo W \bo x, \quad \bo W\in\mathcal S_A. \label{def:z}
\end{align}
We would refer to $\bo W$ as the \emph{collaboration matrix}. It may be noted that, 1) \emph{Particularization:} When $\bo W$ is a diagonal matrix (equivalently, $\bo A$ is the identity matrix $\bo I_N$), our collaborative scheme simplifies to the basic amplify-and-forward strategy \cite{Xiao08},  2) \emph{Collaboration cost:} Any collaboration involving $\bo W\in \mathcal S_A$ is achieved at the expense of (cumulative) power 
\begin{align}
Q_{\bo A}\triangleq \sum_{i=1}^N\sum_{j=1}^N C_{i,j} A_{i,j}, \label{def:PA}
\end{align} 
and 3) \emph{Transmission cost:} The (cumulative) power required for transmission of encoded message $\bo z$ is 
\begin{align}
\begin{split}
P_{\bo W}\triangleq \mathbb E_{\theta,\bo \epsilon}[\bo z^T \bo z]=\tr\left[\bo W \left(\bo \Sigma+\eta^2\bo h\bo h^T\right)\bo W^T\right]. \label{def:PW}
\end{split}
\end{align}


The transformed observations $\bo z$ are assumed to be transmitted to the fusion center through a coherent-MAC channel. In practice, a coherent MAC channel can be realized through \emph{transmit beamforming} \cite{Mudumbai09}, where sensor nodes simultaneously transmit a common message (in our case, all $z_k$-s are scaled versions of a common $\theta$) and the phases of their transmissions are controlled so that the signals constructively combine at the FC. The channel gain at node $n$ is assumed to be $g_n$ and the noise of the coherent-MAC channel $u$ is assumed to be a zero-mean AWGN with variance $\xi^2$. \emph{Perfect knowledge of the channel state $\{g_n\}_{n=1}^N$ and $\xi^2$ is assumed.} Let $\bo g = [g_1,g_2,\ldots,g_N]$.  The output of the coherent-MAC channel (or the input to the fusion center) is 
\begin{align}
y=\bo g^T\bo z+u, \quad u\sim\mathcal N(0,\xi^2). \label{def:y}
\end{align}

Having received $y$, the goal of the fusion center is to obtain an accurate estimate $\widehat \theta$ of the original random parameter $\theta$. We consider the mean square error (MSE) as the distortion metric $D_{\bo W} \triangleq \mathbb E_{\theta,\bo \epsilon, u}\left[(\theta-\widehat\theta)^2; \bo W \right]$. Since the measurement model is (conditionally) linear and Gaussian (see \eqref{def:x}, \eqref{def:z} and \eqref{def:y}), 
\begin{align}
y|\theta\sim \mathcal N(\bo g^T \bo W \bo h \theta, \bo g^T \bo W \bo \Sigma \bo W^T \bo g +\xi^2),
\end{align}
the minimum mean square estimator (MMSE) \cite{Kay93}, $\widehat \theta \triangleq\mathbb E_{\theta,\bo \epsilon, u} [\theta|y]$ is used as the optimum fusion rule. It is well known that MMSE attains the posterior Cram\'er-Rao lower bound,
\begin{align}
D_{\bo W}=\left[\frac{1}{\eta^2} + J_{\bo W}\right]^{-1},  \quad J_{\bo W} \triangleq \frac{(\bo g^T \bo W \bo h)^2}{\bo g^T \bo W \bo \Sigma \bo W^T \bo g +\xi^2},  \label{def:DJW}
\end{align}
where $J_{\bo W}$ denotes the (conditional) Fisher information. It may be noted here that, for the centralized case, i.e., where all the observations $\bo x$ are directly available at the FC, the benchmark performance is,
\begin{align}
D_0\triangleq\left[\frac{1}{\eta^2} + J_0\right]^{-1},  \quad J_0 \triangleq \bo h^T \bo \Sigma^{-1} \bo h.  \label{def:DJ0}
\end{align}

The design of the the collaboration matrix $\bo W$ is critical since it affects both the power requirements and estimation performance of the entire application. Specifically, the following quantities depend on $\bo W$, 1) the resources required to collaborate, $Q_{\text{\textsf{nz}}(\bo W)}$\footnote{Definition of operators $\text{\textsf{nz}}(\cdot)$, $\text{\textsf{zero}}(\cdot)$, and $\text{\textsf{nnz}}(\cdot)$: The operator $\text{\textsf{nz}}:\mathbb R^{N\times N}\rightarrow \{0,1\}^{N\times N}$ is used to specify the non-zero elements of a matrix. If $W_{ij}\neq 0$, then $\left[\text{\textsf{nz}}( \bo W)\right]_{ij}=1$, else $\left[\text{\textsf{nz}} (\bo W)\right]_{ij}=0$. Similarly, the operator $\text{\textsf{zero}}:\mathbb R^{N\times N}\rightarrow \{0,1\}^{N\times N}$ is used to specify the zero elements of a matrix, $\left[\text{\textsf{zero}}( \bo W)\right]_{ij}=1-\left[\text{\textsf{nz}}( \bo W)\right]_{ij}$. The operator $\text{\textsf{nnz}}:\mathbb R^{N\times N}\rightarrow \mathbb Z_{+}$ is used to specify the number of non-zero elements of a matrix.}, as described in \eqref{def:PA}, 2) the resources required to transmit, $P_{\bo W}$, as described in \eqref{def:PW} and 3) the final distortion of the estimate at the FC, $D_{\bo W}$, provided by \eqref{def:DJW}. In this paper, we address the problem of designing the optimum collaboration matrix subject to a (cumulative) power constraint,
\begin{equation}
 \underset{\bo W}{\text{minimize}} \mbox{ } D_{\bo W}, \quad
                           \text{subject to} \mbox{ } P_{\bo W}+Q_{\text{\textsf{nz}}(\bo W)}\le P,
\label{prob:gen}
\end{equation}
where $P$ denotes the (cumulative) power available in the network. It should be noted that, in addition to a cumulative power constraint, there may be individual power constraints corresponding to the various sensor nodes. However, we do not address the individual power constraints in this paper and this issue remains a worthy topic for future research.

Problem \eqref{prob:gen}, in general, has no known globally optimal solution. However, \emph{for the special case when the entries of the collaboration cost matrix $\bo C$ are either zero or infinity, $C_{ij}\in\{0,\infty\}$, we will show that there exists a unique solution for which a closed-form solution can be derived}. Physically, this special case corresponds to the situation when the topology of a network is fixed (and hence not subject to design) and communication among neighbors are relatively inexpensive compared to communication with the FC. Let $\bo A=\text{\textsf{zero}}(\bo C)$ denote the permitted adjacency matrix for such a  situation. Hence, the collaboration cost vanishes, $Q_{\bo A}=0$, and problem \eqref{prob:gen} simplifies to,
\begin{equation}
\underset{\bo W\in \mathcal S_{\bo A}}{\text{minimize}}\mbox{ }D_{\bo W}, \quad \text{subject to}\mbox{ } P_{\bo W}\le P,
 \label{prob:ideal}
\end{equation}
which is an optimization problem in $\text{\textsf{nnz}}(\bo A)$ variables.

Since problem \eqref{prob:ideal} arises out of the assumption of zero-cost for collaboration, we would refer to \eqref{prob:ideal} as the \emph{ideal-collaborative power-allocation} problem. As regards the more general case (problem \eqref{prob:gen} for arbitrary costs $\bo C$ and the topology being subject to design), one can start from the distributed topology $\bo A=\bo I$, and follow a greedy algorithm and augment the collaborative topology with the most power-efficient link at each iteration. This extension is not discussed in this paper and is relegated to a later version of this paper.

\section{Ideal-collaborative power-allocation} \label{sec:ideal}
From \eqref{def:DJW}, we note that minimizing the distortion $D_{\bo W}$ is equivalent to maximizing the (conditional) Fisher information $J_{\bo W}$. Hence problem \eqref{prob:ideal} is equivalent to,
\begin{equation}
\underset{\bo W\in \mathcal S_{\bo A}}{\text{maximize}}\mbox{ }J_{\bo W}, \quad \text{subject to} \mbox{ } P_{\bo W}\le P.
\label{prob:ideal:J}
\end{equation}
Since multiplying $\bo W$ by a scalar $\alpha>1$ (strictly) increases both $J_{\bo W}$ and  $P_{\bo W}$ (and for $\alpha<1$, strictly decreases them), problem \eqref{prob:ideal:J} is equivalent to its converse formulation, where power is minimized subject to a minimum (conditional) Fisher information $J\in(0,J_0)$,
\begin{equation}
\underset{\bo W\in \mathcal S_{\bo A}}{\text{minimize}} \mbox{ } P_{\bo W}, \quad   \text{subject to} \mbox{ } J_{\bo W}\ge J,
 \label{prob:ideal:J:converse}
\end{equation}
in the sense that the optimal solutions $J_{\text{opt}}(P)$ (of \eqref{prob:ideal:J}) and $P_{\text{opt}}(J)$ (of \eqref{prob:ideal:J:converse}) are inverses of one another. Moreover, the optimal solutions hold with active constraints (satisfying equalities). From \eqref{def:PW} and \eqref{def:DJW}, problem \eqref{prob:ideal:J:converse} is further equivalent to,
\begin{equation}
\begin{aligned}
& \underset{\bo W\in \mathcal S_{\bo A}}{\text{minimize}} & & \tr\left[\bo W \left(\bo \Sigma+\eta^2\bo h\bo h^T\right)\bo W^T\right] \\
&                                                             \text{subject to} & & \bo g^T \bo W \left(J \bo \Sigma-\bo h \bo h^T\right) \bo W^T \bo g +J \xi^2 \le 0,
\end{aligned} \label{prob:ideal:implicit}
\end{equation}
which, on closer look, is a quadratically constrained quadratic program (QCQP) in $L\triangleq\text{\textsf{nnz}}(\bo A)$ variables. 

\begin{figure}[t]
\begin{center}
    \includegraphics[width=\figszb \columnwidth]{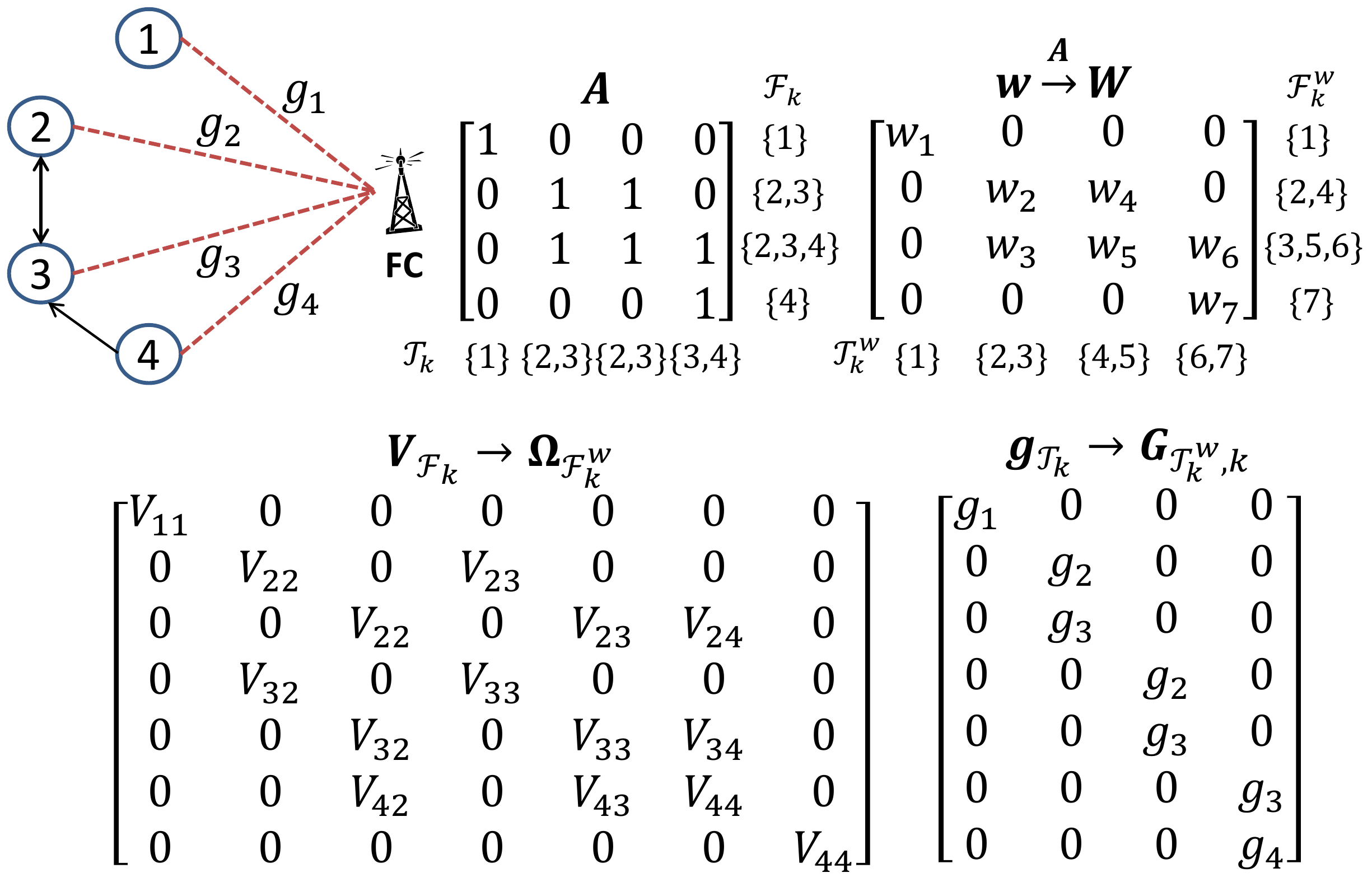}
\vspace{-0.2in}
  \caption{Transformations for QCQP formulation in explicit form - an example.  }
  \label{fig:tx:example}
\end{center}
\vspace{-0.3in}
\end{figure}

An explicit form of QCQP can be obtained from problem \eqref{prob:ideal:implicit} by concatenating the elements of $\bo W$ (column-wise, only those that are allowed to be non-zero), in $\bo w=[w_1,w_2,\ldots,w_L]^T$, and accordingly transforming other constants, 
\begin{align}
\begin{aligned}
& \underset{\bo w}{\text{minimize}} & & \bo w^T \bo \Omega \bo w \\
&                            \text{subject to} & & \bo w^T \bo G \bo Z \bo G^T \bo w  +J \xi^2 \le 0, \mbox{ where }
\end{aligned} \label{prob:ideal:explicit}
\end{align}
\vspace{-.15in}
\begin{align}
\begin{split}
\bo w  \stackrel{\bo A}{\rightarrow} \bo W, \bo V  \stackrel{\bo A}{\rightarrow} \bo \Omega,  \bo g  \stackrel{\bo A}{\rightarrow} \bo G, \mbox{ and } \\
\bo V \triangleq \bo \Sigma+\eta^2\bo h\bo h^T, \bo Z \triangleq  J \bo \Sigma-\bo h \bo h^T.
\end{split} \label{def:VZ}
\end{align}
We illustrate the relevant  transformations through an example, in Figure \ref{fig:tx:example}, with $N=4$ nodes and $3$ collaborating links, i.e., total $L=7$ non-zero coefficients. Based on topology $\bo A$, node $k$ sends its observations \emph{to} nodes $\mathcal T_k$ and receives observations \emph{from} nodes $\mathcal F_k$. The notations $\mathcal T_k^w$ and $\mathcal F_k^w$ similarly denote the respective indices of $\bo w$ as obtained from $\bo W$. The matrix $\bo \Omega$ is formed from $\bo V$ by copying the (sub)matrices $\bo V_{\mathcal F_k}\rightarrow \bo \Omega_{\mathcal F_k^w}$ for $k=1,2,\ldots,N$, satisfying $\tr(\bo W\bo V\bo W^T)=\bo w^T\bo \Omega \bo w$. The matrix $\bo G$ is similarly formed from vector $\bo g$ by copying the elements $\bo g_{\mathcal T_k}\rightarrow \bo G_{\mathcal T_k^w,k}$ for all $k$,  satisfying $\bo g^T \bo W=\bo w^T \bo G$.

The solution to problem \eqref{prob:ideal:explicit} (equivalently, problems \eqref{prob:ideal:J}, \eqref{prob:ideal:J:converse} and \eqref{prob:ideal:implicit}) is summarized in Theorem \ref{theorem:ideal:QCQP:lbl}.
\begin{theorem:ideal:QCQP} \emph{(Power-Distortion tradeoff for Linear Coherent Ideal-Collaborative Estimation)} \label{theorem:ideal:QCQP:lbl}
Assuming $\bo \Sigma$ to be positive definite, the tradeoff between (conditional) Fisher Information and (cumulative) transmission power is
\begin{align}
\begin{split}
J_{\text{opt}}(P)=\bo h^T\left(\bo \Sigma+\bo \Gamma/P_\xi  \right)^{-1} \bo h, \mbox{ where} \\
P_\xi \triangleq P/\xi^2, \bo \Gamma \triangleq \left(\bo G^T \bo \Omega^{-1}\bo G\right)^{-1},
\end{split} \label{theorem:ideal:tradeoff}
\end{align}
which is achieved when the weights of collaboration matrix are
\begin{align}
\bo w_{\text{opt}}=\kappa \bo \Omega^{-1}\bo G \bo \Gamma\left(\bo \Sigma+ \bo \Gamma/P_\xi  \right)^{-1} \bo h, \label{theorem:ideal:weights}
\end{align}
where the scalar $\kappa$ is such that $\bo w_{\text{opt}}^T\bo \Omega \bo w_{\text{opt}}=P$. Equivalently, for $J\in(0,J_0)$, $P_{\text{opt}}(J)=J\xi^2\mu_+(J)$, where $\mu_+(J)$ is the only positive solution to the generalized eigenvalue problem $(\bo \Gamma+\mu \bo Z)\bo v=0$ (note that $\bo Z$ is a function of $J$).
\end{theorem:ideal:QCQP}

\begin{IEEEproof} See Appendix \ref{app:ideal:QCQP}.  \end{IEEEproof}

\begin{figure}
\vspace{-0.1in}
\begin{center}
    \includegraphics[width=\figszc \columnwidth]{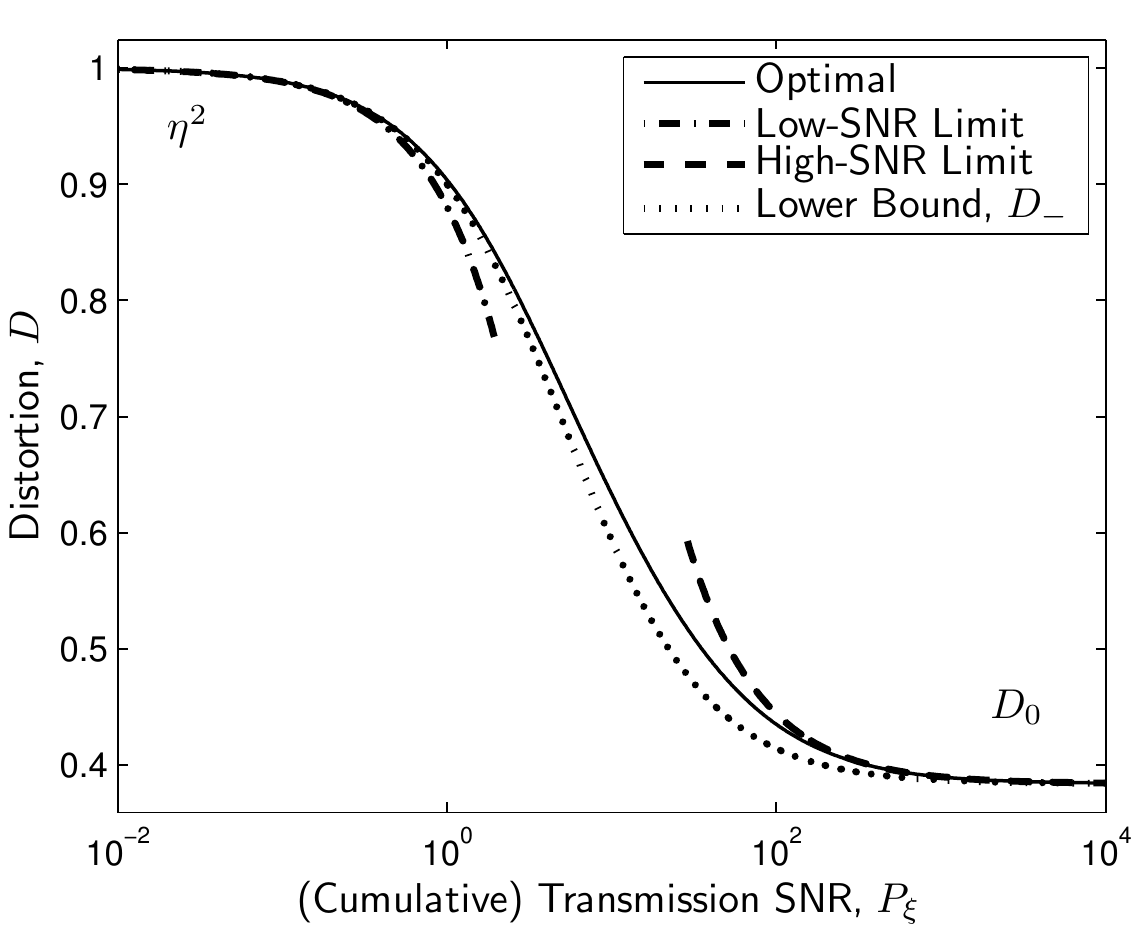}
\vspace{-0.1in}
  \caption{Power-distortion tradeoff from Theorem \ref{theorem:ideal:QCQP:lbl}.}
  \label{fig:cooperative:snr}
  \end{center}
\vspace{-0.3in}
\end{figure}

Theorem \ref{theorem:ideal:QCQP:lbl} is important since it shows the effect of (cumulative) transmit power and the topology on the estimation performance. Corresponding to the example topology in Figure \ref{fig:tx:example} and randomly chosen system parameters $\bo h, \bo \Sigma$ and $\bo g$, a typical power-distortion tradeoff curve is shown in Figure \ref{fig:cooperative:snr} (bold line). Some remarks regarding Theorem \ref{theorem:ideal:QCQP:lbl} are in order.

\emph{Remark 1 (Distributed and fully connected cases):} For the distributed scenario, $\bo A=\bo I$, and we have $\bo w=\text{\textsf{diag}}(\bo W)$,\footnote{Definition of operators $\text{\textsf{diag}}(\cdot)$ and  $\text{\textsf{vec}}(\cdot)$: While operating on a matrix, $\text{\textsf{diag}}:\mathbb R^{N\times N}\rightarrow \mathbb R^N$ is used to extract the diagonal elements. While operating on a vector, $\text{\textsf{diag}}:\mathbb \mathbb R^N\rightarrow R^{N\times N}$ is used to construct a matrix by specifying only the diagonal elements, the other elements being zero. The vectorization operator $\text{\textsf{vec}}:\mathbb R^{N\times N}\rightarrow \mathbb R^{N^2}$ stacks up all the elements of a matrix column-by-column. } 
$\bo \Omega=\text{\textsf{diag}}(\text{\textsf{diag}}(\bo V))$ and $\bo G=\text{\textsf{diag}}(\bo g)$. Furthermore, when $\bo \Sigma$ is diagonal (equivalently, when observation noise is spatially uncorrelated), equation \eqref{theorem:ideal:tradeoff} reduces to 
\begin{align}
J_{\text{opt}}^\text{dist}(P)=\sum_{n=1}^N \frac{h_n^2}{\sigma_n^2+\frac{\sigma_n^2+\eta^2 h_n^2}{P_\xi g_n^2}}, \mbox{ where } \sigma_n^2\triangleq \Sigma_{n,n},
\end{align} 
precisely the result obtained in \cite{Xiao08}. 

For the fully connected scenario, $\bo A=\bo 1\bo 1^T$, we have $\bo w=\text{\textsf{vec}}(\bo W)$, $\bo \Omega=\bo V \otimes \bo I$, $\bo G=\bo I \otimes \bo g$, and subsequently the following result.

\begin{theorem:conn} \emph{(Power-distortion tradeoff for fully connected topology):} \label{theorem:conn:lbl}
\begin{align}
J_\text{opt}^\text{conn}(P)=\left[\frac{1}{J_0}+\frac{\eta^2+\frac{1}{J_0}}{P_\xi \|\bo g\|^2} \right]^{-1}, \bo W_\text{opt}^\text{conn}\propto \bo g \bo h^T\bo \Sigma^{-1}. \label{J:conn}
\end{align}
Furthermore, distortion resulting from \eqref{J:conn} is information theoretically optimal.
\end{theorem:conn}

\begin{IEEEproof} See Appendix \ref{app:conn}.  \end{IEEEproof}

The information theoretic optimality is expected since a fully connected network is equivalent to the centralized scenario with effective channel gain $\|\bo g\|$.

\begin{table}
\begin{center}
\begin{tabular}{|c|c|c|}
  \hline
         	& Low SNR, $\lim_{P_\xi\rightarrow 0}$               			&  High SNR, $\lim_{P_\xi\rightarrow \infty}$ \\
  \hline
   $J	$	& $P_\xi \bo h^T\bo \Gamma^{-1} \bo h$ 					& $J_0-P_{\xi}^{-1} \bo h^T\bo \Sigma^{-1}\bo \Gamma \bo \Sigma^{-1} \bo h$\\ 
  \hline
   $D$	& $\eta^2-\eta^4 P_\xi \bo h^T\bo \Gamma^{-1} \bo h$ 	& $D_0+D_0^2P_{\xi}^{-1} \bo h^T\bo \Sigma^{-1}\bo \Gamma \bo \Sigma^{-1} \bo h$\\
  \hline
 $\bo w$& $\kappa P_\xi \bo \Omega^{-1} \bo G \bo h$				& $\kappa \bo \Omega^{-1} \bo G \bo \Gamma \bo \Sigma^{-1} \bo h$ \\
  \hline
\end{tabular}
\end{center} 
\caption{Distortion and optimal weights for low and high SNR limits.} \label{tbl:ideal:snr}
\vspace{-0.3in}
\end{table}

\emph{Remark 2 (Limits and a lower bound):}
For better understanding the dependence of distortion $D$ on (cumulative) SNR $P_\xi$, we compute the low and high SNR limits of distortion (and optimal weights, upto second order Taylor series) in Table \ref{tbl:ideal:snr}. For any topology $\bo A$ (and consequently $\bo \Gamma$), provided a large (cumulative) power is available, the resultant distortion approaches that of the centralized case, $D_0$ (see \eqref{def:DJ0}). 
In low-SNR situations, the distortion approaches that of the prior, $\eta^2$. 
Towards the goal of obtaining a simpler approximation of \eqref{theorem:ideal:tradeoff} for both the low and high SNR regimes, we obtain the following result.

\begin{theorem:lowerbd}[Lower bound on distortion] \label{theorem:lowerbd:lbl}
Define,
\begin{align}
J_{+}(P)\triangleq\left[\frac{1}{J_0}+\frac{1}{P_\xi \bo h^T \bo\Gamma^{-1}\bo h}\right]^{-1}, D_{-}(P)\triangleq \left[\frac{1}{\eta^2}+J_{+}(P)\right]^{-1}.  \label{lowerbd}
\end{align} 
Then, $J_{+}(P)\ge J_{\text{opt}}(P)$ and hence $D_{-}(P)\le D_{\text{opt}}(P)$.
\end{theorem:lowerbd}

\begin{IEEEproof} Follows from equation \eqref{theorem:ideal:tradeoff} and the fact that both $\bo \Sigma$ and (consequently) $\bo \Gamma$ are positive definite, then applying Lemma \ref{theorem:inequality:lbl} (Appendix \ref{app:inequality}). \end{IEEEproof}

Both the high and low-SNR limits and the lower bound $D_{-}$ are displayed in Figure \ref{fig:cooperative:snr}. From Figure \ref{fig:cooperative:snr}, we verify that both the low and high SNR limits are quite accurate (in their respective regimes) and the lower bound, while accurate in both the limits, always satisfy $D_{-}<D$. 

\emph{Remark 3 (Decentralized computation of collaborative strategies):} The optimal combining weights in Table \ref{tbl:ideal:snr}, besides being accurate in the low and high-SNR regimes respectively, have appealing interpretations that can facilitate decentralized computation of collaborative strategies, thereby requiring lesser coordination with the fusion center and facilitating faster adaptation to dynamically changing topologies. Firstly, it can be shown that, $\bo w \propto \bo \Omega^{-1} \bo G \bo h$ (low-SNR regime) corresponds to the case where each node is performing local-MMSE estimation. Computation of the optimal combining weights can hence be performed from local observation and covariance models only. Secondly, $\bo w \propto  \bo \Omega^{-1} \bo G \bo \Gamma \bo \Sigma^{-1} \bo h$ (high-SNR regime) can be shown to correspond to the solution of a convex linearly constrained quadratic program (LCQP) with separable objective function, which can be efficiently solved in a decentralized manner \cite{Boyd11}. We relegate the details to a future possible extension of this paper.

\begin{figure}[b]
\vspace{-0.2in}
\begin{center}
    \includegraphics[width=\figszd \columnwidth]{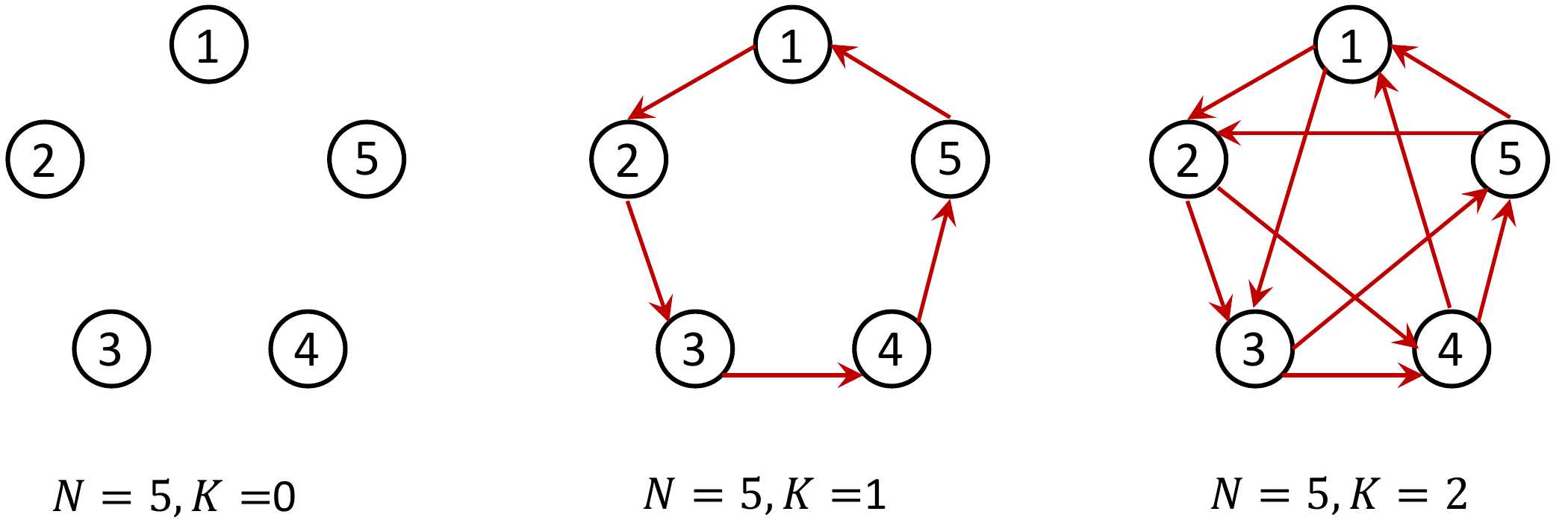}
\vspace{-0.1in}
  \caption{Directed cycle graphs, $K$-connected.}
  \label{fig:cooperative:cycle}
\vspace{-0.1in}
  \end{center}

\end{figure}

\emph{Remark 4 (Closed form results for regular graphs): } For some combinations of signal parameters, network topology and channel gains, the power-distortion tradeoff can be explicitly derived. In Figure \ref{fig:cooperative:cycle}, we display a class of graphs, namely the $K$-connected directed cycle, in which each node shares its observations with the next $K$ nodes. Note that $K=0$ denotes the distributed scenario while $K=N-1$ denotes the fully connected scenario.

\begin{theorem:cycle} \emph{(Homogeneous and equicorrelated sensor network with cycle topology)} \label{theorem:cycle:lbl}
Assume a collaborative sensor network with, 1) identical observation gains, $\bo h=h_0 \bo 1$, 2)  equicorrelated and homogeneous observation noise, $\bo \Sigma=\sigma^2((1-\rho)\bo I+\rho \bo 1\bo 1^T)$, where $\rho\in[0,1)$,   3) $K$-connected directed cycle as the neighborhood adjacency matrix $\bo A$, and 4) identical channel gains, $\bo g=g_0 \bo 1$. For such a problem setup, the lower bound in \eqref{lowerbd} is actually an equality, i.e., $D_{\text{opt}}(P)=D_{-}(P)$, with $\bo W_\text{opt}^{\mathcal C(K)}(P)\propto \bo A$ and
\begin{align}
\small{J_\text{opt}^{\mathcal C(K)}(P)=\left[\frac{1}{J_0}+\frac{1}{P_\xi N g_0^2}\left\{\eta^2+\frac{\sigma^2}{h_0^2}\left(\rho+\frac{1-\rho}{K+1}\right) \right\} \right]^{-1}}. \label{J:cycle}
\end{align}
\end{theorem:cycle}

\begin{IEEEproof} See Appendix \ref{app:cycle}. \end{IEEEproof}

From Proposition \ref{theorem:cycle:lbl}, we readily infer the conditions under which collaboration can be beneficial. For $K=0,1,\ldots,N-1$, let us denote by $P_{\text{opt}}^{\mathcal C(K)}(J)$ the (minimum) power required to obtain some prespecified distortion $D$ ($J$ and $D$ are related by \eqref{def:DJW}). Then the (relative) power (RPS) savings  obtained due to collaboration is (from \eqref{J:cycle}),
\begin{align}
\text{RPS}(K,J)\triangleq1-\frac{P_{\text{opt}}^{\mathcal C(K)}(J)}{P_{\text{opt}}^{\mathcal C(0)}(J)}=\frac{(1-\rho)(1-\frac{1}{K+1})}{1+\frac{\eta^2h_0^2}{\sigma^2}},
\end{align} 
which represents the gain compared to distributed scenario $(K=0)$. Firstly, we note that $\text{RPS}(K,J)\in [0,1)$ (since $\rho\in[0,1)$ and $K\ge 0$), which shows that it is always beneficial to collaborate, assuming cost-free collaboration. Also, more (relative) power is saved when, 1) the collaboration among nodes increases (higher $K$), 2) the observation noise is less correlated (lower $\rho$), and 3) the local-SNR is small (smaller $\gamma\triangleq\frac{\eta^2h_0^2}{\sigma^2}$). When local-SNR is large, say $\gamma=100$, then even a fully connected network can provide only a power saving of $1\%$. On the other hand, if the local-SNR is small, say $\gamma=1$, then a fully connected network can provide upto $50\%$ power savings.

\begin{figure}[t]
\begin{center}
    \includegraphics[width=\figsze \columnwidth]{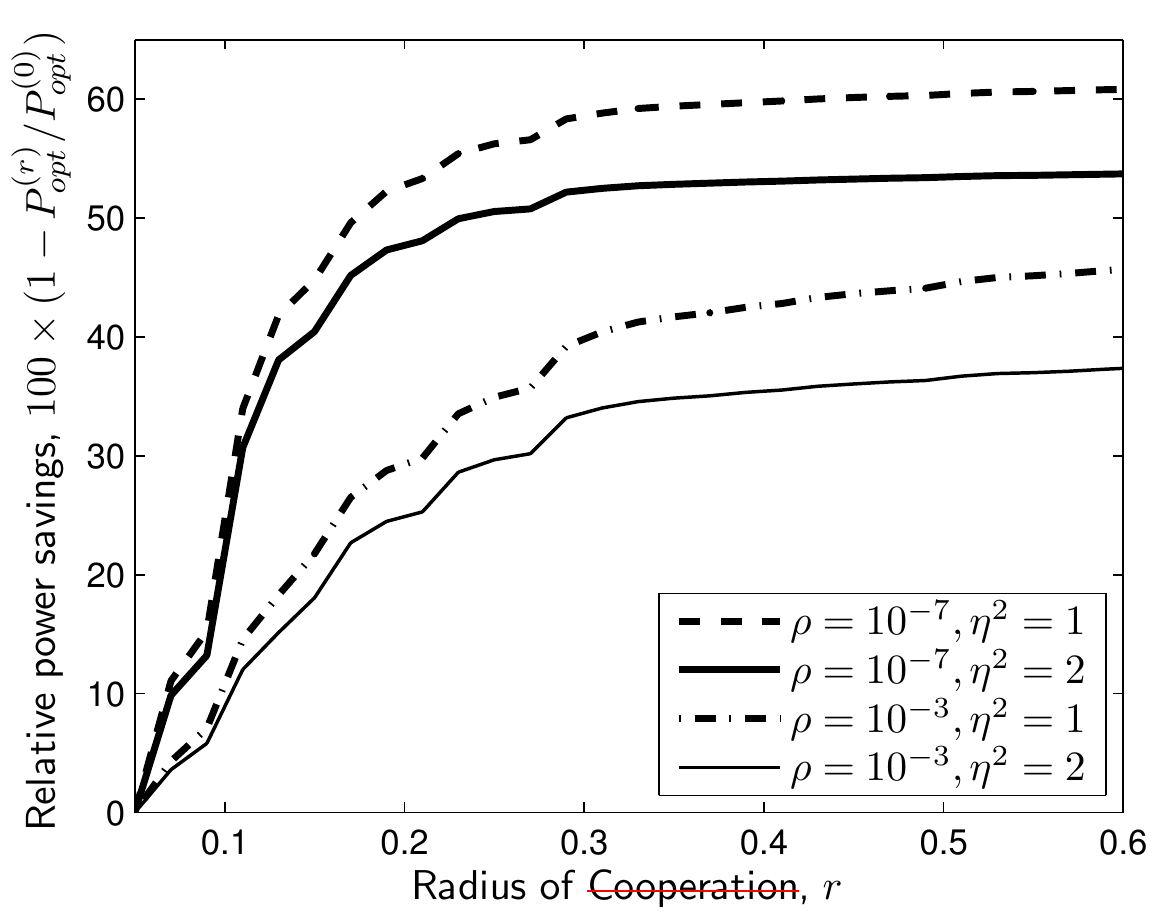} 
\vspace{-0.25in}
  \caption{Efficiency in power achieved through collaboration in a 50-node random geometric graph. }
  \label{fig:cooperative:powereff}
  \end{center}
\vspace{-0.3in}
\end{figure}

\section{Numerical Results} \label{sec:numerical}
To demonstrate the (cumulative) power saved due to collaboration and to investigate whether the insights obtained from Proposition \ref{theorem:cycle:lbl} extend to more complicated scenarios, we consider the following simulation setup. The spatial placement and neighborhood structure is modeled as a Random Geometric Graph, $\text{RGG}(N,r)$ \cite{Freris10}, where sensors are uniformly distributed over a unit square with bidirectional communication links present only for pairwise distances at most $r$, i.e., $\bo A$ such that $A_{i,j}=\indicator{d_{i,j}\le r}$.
The noise is modeled as a homogeneous and exponentially correlated Gaussian covariance matrix, i.e., $\bo \Sigma$ is such that $\Sigma_{i,j}=\sigma^2 \rho^{d_{i,j}}$, where $\rho\in(0,1)$ is indicative of the degree of spatial correlation. A smaller value of $\rho$ indicates lower correlation with $\rho\rightarrow 0$ signifying completely independent observations. Specifically, we consider $\rho=10^{-3}$ and $\rho=10^{-7}$ to contrast the effect of correlation (for sensor nodes apart by distance $d_{i,j}=0.1$, the actual correlations are $\rho^{0.1}\approx0.5$ and $\rho^{0.1}\approx0.2$ respectively). We consider $N=50$ nodes with identical local-SNR (specifically, $\sigma^2=0.5$, $\bo h=\bo 1$ with $\eta^2=1$ and $\eta^2=2$ for two separate runs). The individual channel gains were generated by uniform random numbers in the range $(0,1]$. At each instance, power was allocated to satisfy the  pre-specified distortion performance of $\frac{\eta^2+D_0}{2}$.  We display the power savings obtained after collaborating through $\text{RGG}(N,r)$ topology, $P_\text{opt}^{(0)}-P_\text{opt}^{(r)}$, as a percentage of the power required for the distributed case, $P_\text{opt}^{(0)}$, for increasing radius of collaboration $r$, in Figure \ref{fig:cooperative:powereff}. We note that significant power is saved through collaboration for different magnitudes of local-SNR, $\eta^2$, and varying degrees of spatial correlation, $\rho$. Also, we observe that (relative) power savings seem to increase with lower spatial correlation and lower local-SNR, which were also the insights obtained from the simpler example considered in Proposition \ref{theorem:cycle:lbl}.

\section{Conclusion}
In this paper, we addressed the problem of collaborative estimation in a sensor network where sensors communicate with the FC using a coherent MAC channel. For the scenario when the collaborative topology is fixed and collaboration is cost-free, we obtained the optimal power-distortion tradeoff in closed-form by solving a QCQP problem. Through the use of both theoretical and numerical results, we established that collaboration helps to substantially lower the power requirements in a network, specially in low local-SNR scenario. As future work, we wish to explore the collaborative estimation problem when the parameter to be estimated is a vector with correlated elements. The issue of collaboration with non-zero cost, as mentioned earlier, is also important. Finally, collaboration in the presence of individual power constraints (in addition to cumulative) is another topic worthy of future research.



\appendices
\section{Proof of Theorem \ref{theorem:ideal:QCQP:lbl}} \label{app:ideal:QCQP}
Note in \eqref{prob:ideal:explicit} that, though $\bo \Omega$ is positive definite (since $\bo \Sigma$ is), $\bo Z$ is not (in fact, $\bo Z$ has exactly one negative eigenvalue), and hence problem \eqref{prob:ideal:explicit} is not convex. However, a QCQP with exactly one constraint (as in problem \eqref{prob:ideal:explicit}) still satisfies strong duality\footnote{The technical requirement of Slatar's constraint qualification is not discussed, but can be shown to be satisfied} (e.g., Appendix B, \cite{Boyd04}) and hence the optimal solution to \eqref{prob:ideal:explicit} satisfies the Karush-Kuhn-Tucker (KKT) conditions. Therefore, for some $\mu>0$,
\begin{align}
&(\bo \Omega +\mu \bo G \bo Z \bo G^T)\bo w_{\text{opt}}=\bo 0_L, (\mbox{KKT, } \eqref{prob:ideal:explicit}) \label{app:kkt} \\
\Leftrightarrow\quad &(\bo I_L+\mu \bo \Omega^{-1}\bo G \bo Z \bo G^T)\bo w_{\text{opt}}=\bo 0_L, (\bo \Omega \mbox{ is full rank}) \label{app:v2w} \\
\Rightarrow\quad &(\bo G^T+\mu \bo G^T\bo \Omega^{-1}\bo G \bo Z \bo G^T)\bo w_{\text{opt}}=\bo 0_N \nonumber \\
\Leftrightarrow\quad &(\bo I_N+\mu \bo G^T\bo \Omega^{-1}\bo G \bo Z) \bo v=\bo 0_N, \bo v\triangleq \bo G^T\bo w_{\text{opt}} \nonumber \\
\Leftrightarrow\quad &(\bo \Gamma+\mu \bo Z) \bo v=\bo 0_N, ( \bo \Gamma \mbox{ is full rank},  \eqref{theorem:ideal:tradeoff}) \label{app:GZv} \\
\Leftrightarrow\quad &(\bo \Gamma+\mu  J \bo \Sigma-\mu \bo h \bo h^T) \bo v=\bo 0_N, (\mbox{definition of }\bo Z, \eqref{def:VZ}) \nonumber \\
\Leftrightarrow\quad &(\bo I_N-\mu (\bo \Gamma+\mu  J \bo \Sigma)^{-1}\bo h \bo h^T) \bo v=\bo 0_N  \label{app:eigvec} \\
\Rightarrow\quad &(1-\mu \bo h^T (\bo \Gamma+\mu  J \bo \Sigma)^{-1}\bo h) (\bo h^T\bo v)=0, (\mbox{multiplied by } \bo h^T) \nonumber \\
\Rightarrow\quad &f(\mu)=0,  (\mbox{define } f(\mu)\triangleq1-\mu \bo h^T (\bo \Gamma+\mu  J \bo \Sigma)^{-1}\bo h)  \label{app:eigval}
\end{align}
where \eqref{app:eigval} is because $\bo h^T\bo v=\bo h^T \bo G^T \bo w_{\text{opt}}$ is the numerator of Fisher information in \eqref{def:DJW} and hence non-zero. Note that, since $\bo \Gamma$ and $\bo \Sigma$  are both positive definite, $f(\mu)$ is monotonically decreasing for $\mu>0$, with $f(0)=1$ and $f(\infty)\searrow1-J_0/J<0$ (since $J<J_0$). Hence \eqref{app:eigval} must have a unique positive root (denote it as $\mu_+(J)$). Since constraint is active at the optimal solution,
\begin{align}
\bo w_{\text{opt}}^T \bo G \bo Z \bo G^T \bo w_{\text{opt}}  +J \xi^2=0, \label{ideal:active}
\end{align}
which, along with $\bo w_{\text{opt}}^T \bo \Omega \bo w_{\text{opt}}=P$ and \eqref{app:kkt}, leads to $J=P_\xi/\mu_+(J)$. Substituted in \eqref{app:eigval}, this leads to equation \eqref{theorem:ideal:tradeoff}. From \eqref{app:eigvec}, we readily obtain $\bo v\propto (\bo \Gamma+\mu  J \bo \Sigma)^{-1}\bo h$, and from \eqref{app:v2w}, we obtain 
\begin{align*}
\bo w&\propto \bo \Omega^{-1}\bo G \bo Z \bo v \\
&= \bo \Omega^{-1}\bo G \bo \Gamma \bo v, (\mbox{since } \bo Z \bo v\propto \bo \Gamma \bo v, \eqref{app:GZv})
\end{align*}
which, alongwith $\mu_+(J)=P_\xi/J$, gives \eqref{theorem:ideal:weights}.

\section{Proof of Proposition \ref{theorem:conn:lbl}} \label{app:conn}
Note that $\bo \Omega=\bo V \otimes \bo I$ and $\bo G=\bo I \otimes \bo g$. Hence, 
\begin{align}
\bo \Gamma=(\bo G^T\bo \Omega^{-1}\bo G)^{-1}= \frac{\bo V}{\|\bo g\|^2}, \quad (\|\bo g\|^2= \bo g^T\bo g).
\end{align}
Substituting this value of $\bo \Gamma$ in \eqref{theorem:ideal:tradeoff} (recall, $\bo V=\bo \Sigma+\eta^2\bo h\bo h^T$),
\begin{align}
J_{\text{opt}}^{\text{conn}}&=\bo h^T (\bo \Sigma+\bo \Gamma/P_\xi)^{-1}\bo h =\bo h^T (\alpha \bo \Sigma+\beta \bo h\bo h^T)^{-1} \bo h, \label{J:conn:interm}
\end{align}
where $\alpha=1+\frac{1}{P_\xi \|\bo g\|^2}$ and $\beta=\frac{\eta^2}{P_\xi \|\bo g\|^2}$.
From the Woodbury matrix identity, for any $\alpha\neq 0, \beta$, (recall, $J_0=\bo h^T \bo \Sigma^{-1} \bo h$),
\begin{align}
&(\alpha \bo \Sigma+\beta \bo h\bo h^T)^{-1}=\frac{\bo \Sigma^{-1}}{\alpha}-\frac{\beta \bo \Sigma^{-1}\bo h \bo h^T \bo \Sigma^{-1}}{\alpha(\alpha+\beta J_0)} \\
\Rightarrow\quad& (\alpha \bo \Sigma+\beta \bo h\bo h^T)^{-1}\bo h =\frac{\bo \Sigma^{-1}\bo h}{\alpha+\beta J_0}, \label{app:conn:tmp1}
\end{align}
from which $\bo h^T(\alpha \bo \Sigma+\beta \bo h\bo h^T)^{-1}\bo h=\frac{J_0}{\alpha+\beta J_0}$. Applied to \eqref{J:conn:interm}, this yields  the expression of $J_{\text{opt}}^{\text{conn}}$ in \eqref{J:conn}. Also,
\begin{align}
\bo w_{\text{opt}}^{\text{conn}}&\propto \bo \Omega^{-1} \bo G \bo \Gamma (\bo \Sigma+\bo \Gamma/P_\xi)^{-1}\bo h, \quad (\mbox{from  \eqref{theorem:ideal:weights}}) \nonumber \\
&\propto \bo \Omega^{-1} \bo G \bo \Gamma \bo \Sigma^{-1}\bo h, \quad (\mbox{see \eqref{J:conn:interm} and \eqref{app:conn:tmp1}}) \nonumber \\
&\propto \bo \Omega^{-1} \bo G \bo h, \quad (\bo \Gamma\propto \bo V, \bo V\bo \Sigma^{-1}\bo h\propto \bo h) \nonumber \\
&\propto (\bo V^{-1}\bo h) \otimes \bo g, \quad (\bo \Omega^{-1} \bo G=\bo V^{-1}\otimes \bo g) \nonumber \\
&\propto (\bo \Sigma^{-1}\bo h) \otimes \bo g, \quad (\bo V^{-1}\bo h\propto \bo \Sigma^{-1}\bo h, \mbox{ see \eqref{app:conn:tmp1}})
\end{align}
which implies that $\bo W_{\text{opt}}^{\text{conn}}\propto\bo g\bo h^T\bo \Sigma^{-1}$.

From Corollary $2.3.5$ of \cite{Gastpar07chapter}, the sum-rate required to encode a single-dimensional real-valued Gaussian source with variance $\eta^2$, observed through the vector $\bo h$ and Gaussian observation noise with covariance $\bo \Sigma$, in such a way that reconstruction incurs an average distortion of at most $D$, satisfies
\begin{align} 
R_{\text{tot}}\ge \frac{1}{2}\log \frac{\lambda}{D-D_0}, \mbox{ where } \lambda=\frac{\eta^4 J_0}{1+\eta^2 J_0}.
\end{align}
Since, for a fixed sum-power $P$, the sum-rate has to be lesser that the (centralized) capacity of the coherent MAC channel, i.e., $R_{\text{tot}}\le C$, where $C=\frac{1}{2}\log (1+\|\bo g \|^2 P_\xi)$, we obtain
\begin{align} 
1+\|\bo g \|^2 P_\xi\ge  \frac{\eta^4 J_0}{(D-D_0)(1+\eta^2 J_0)}.
\end{align}
Replacing $D$ by $J$ (recall, $J=\frac{1}{D}-\frac{1}{\eta^2}$) and after some algebra, we obtain,
\begin{align} 
J\le J_{\text{opt}}^{\text{conn}},
\end{align}
where $J_{\text{opt}}^{\text{conn}}$ is defined in \eqref{J:conn}. Hence, a fully-connected network that performs cost-free linear collaboration achieves information theoretically optimal performance.

\section{} \label{app:inequality}
\begin{theorem:inequality}[An inequality] \label{theorem:inequality:lbl}
For any $N$-dimensional vector $\bo p$ and $N\times N$ symmetric positive definite matrices $\bo A$ and $\bo B$,
\begin{align}
\frac{1}{\bo p^T \left(\bo A+\bo B\right)^{-1} \bo p} \ge \frac{1}{\bo p^T \bo A^{-1} \bo p}+\frac{1}{\bo p^T \bo B^{-1} \bo p}. \label{posdef:inequality}
\end{align}
\end{theorem:inequality}
\begin{IEEEproof} Since $\bo A,\bo B \in \mathcal S^{++}$, $\bo A^{-\frac{1}{2}}\bo B\bo A^{-\frac{1}{2}}\in \mathcal S^{++}$. Define by $\bo U$ and $\bo \Lambda$ the following eigendecomposition $\bo A^{-\frac{1}{2}}\bo B\bo A^{-\frac{1}{2}}=\bo U\bo \Lambda \bo U^T$. Hence $\lambda_n>0,\forall n$. Define $\bo q=\bo U^T\bo A^{-\frac{1}{2}} \bo h$. Note that 
\begin{align}\begin{split}
\bo q^T \bo q&=\bo p^T \bo A^{-1} \bo p,\\ 
\bo q^T \bo \Lambda^{-1}\bo q&=\bo p^T \bo B^{-1} \bo p, \mbox{ and }\\
\bo q^T (\bo I+\bo \Lambda)^{-1}\bo q&=\bo p^T \left(\bo A+\bo B\right)^{-1} \bo p.
\end{split}\end{align}
Hence, to prove \eqref{posdef:inequality}, it suffices to show that
\begin{align*}
\frac{1}{\sum_{n=1}^N\frac{q_n^2}{1+\lambda_n}} \ge \frac{1}{\sum_{n=1}^N q_n^2}+\frac{1}{\sum_{n=1}^N\frac{q_n^2}{\lambda_n}},
\end{align*}
or equivalently, with $a_n\triangleq \frac{1}{1+\lambda_n}$ and $b_n\triangleq\frac{1+\lambda_n}{\lambda_n}$,
\begin{align}
\sum_{n=1}^N q_n^2 \sum_{n=1}^N q_n^2 a_n b_n &\ge  \sum_{n=1}^N q_n^2 a_n \sum_{n=1}^N q_n^2 b_n. \label{chebyshev}
\end{align}
Since $\lambda_n>0,\forall n$, both $a_n$ and $b_n$ are decreasing functions of $\lambda_n$. Hence inequality \eqref{chebyshev} follows from the Chebyshev's (sum) inequality (page 240, equation 1.4, \cite{Mitrinovic93}). Equality holds if and only if, for all indices $k$ for which $q_k\neq 0$ (denote such a set by $\text{\textsf{ixnz}}(\bo q)$), the eigenvalues are similar. That is, iff $\lambda_k=\lambda$, $\forall k\in\text{\textsf{ixnz}}(\bo q)$.
\end{IEEEproof}

\section{Proof of Proposition \ref{theorem:cycle:lbl}} \label{app:cycle}
To show $D_{\text{opt}}(P)=D_-(P)$, we can show that the condition for equality in Lemma \ref{theorem:inequality:lbl} is satisfied. However, we provide a simpler proof. First we will show that $\bo h$ is an eigenvector of both $\bo \Sigma$ and $\bo \Gamma$, i.e., $\bo \Sigma\bo h=\lambda\bo h$ and $\bo \Gamma\bo h=\mu \bo h$, where the eigenvalues $\lambda$ and $\mu$ will be derived later. Therefore $(\bo \Sigma+\bo \Gamma/P_\xi)\bo h=(\lambda+\mu/P_\xi)\bo h$ and hence, from \eqref{theorem:ideal:tradeoff},
\begin{align}
J_{\text{opt}}^{\mathcal C(k)}&=\bo h^T (\bo \Sigma+\bo \Gamma/P_\xi)^{-1}\bo h =\frac{h_0^2 N}{(\lambda+\mu/P_\xi)}. \label{J:eig}
\end{align}
We next find $\lambda$ and $\mu$. From definitions of $\bo \Sigma$ and  $\bo h$, we directly have 
\begin{align}
\lambda=\sigma^2((1-\rho)+\rho N). \label{eig:lambda}
\end{align}
so that $J_0=\bo h^T\bo \Sigma^{-1}\bo h=\frac{h_0^2 N}{\lambda}$. Let $\widetilde{K}\triangleq K+1$. From the transformation $\bo V  \stackrel{\bo A}{\rightarrow} \bo \Omega$ in \eqref{def:VZ}, we note that $\bo \Omega$ consists of $N$ blocks of identical $\widetilde{K}\times \widetilde{K}$ sub-matrices. For $k=1,2,\ldots,N$,
\begin{align} 
\bo \Omega_{\mathcal F_k^w}&=\bo \Sigma_{\mathcal F_k}+\eta^2 h_0^2 \bo 1_{\widetilde{K}}\bo 1_{\widetilde{K}}^T =\alpha \bo I_{\widetilde{K}}+\beta \bo 1_{\widetilde{K}}\bo 1_{\widetilde{K}}^T, 
\end{align}
where $\alpha=\sigma^2(1-\rho)$ and $\beta=\sigma^2\rho+\eta^2 h_0^2$.  Hence $\bo \Omega \bo 1_L=(\alpha+\beta \widetilde{K}) \bo 1_L$ and therefore $\bo \Omega^{-1} \bo 1_L=\frac{1}{\alpha+\beta \widetilde{K}} \bo 1_L$.  Similarly, from the transformation $\bo g  \stackrel{\bo A}{\rightarrow} \bo G$ in \eqref{def:VZ}, $\bo G$ consists of columns $\bo G_{\mathcal T_k^w,k}=g_0 \bo 1_{\widetilde{K}}$. Hence $\bo G \bo 1_N=g_0 \bo 1_L$ and $\bo G^T \bo 1_L=g_0 \widetilde{K} \bo 1_N$. Next, $\mu$ is obtained by inverting the eigenvalue of $\bo \Gamma^{-1} \bo h$,
\begin{align}
\bo \Gamma^{-1} \bo h&= h_0 \bo G^T \bo \Omega^{-1} \bo G \bo 1_N  \nonumber \\
&=h_0 g_0 \bo G^T \bo \Omega^{-1} \bo 1_L \nonumber \\
&=\frac{h_0 g_0}{\alpha+\beta \widetilde{K}} \bo G^T \bo 1_L \nonumber \\
&=\frac{h_0 g_0^2 \widetilde{K}}{\alpha+\beta \widetilde{K}} \bo 1_N,\mbox{ } \left(\rightarrow\frac{1}{\mu} \bo h, \Rightarrow \mu=\frac{\alpha+\beta \widetilde{K}}{g_0^2 \widetilde{K}}\right). \label{eig:mu}
\end{align}
From \eqref{J:eig}, \eqref{eig:lambda} and \eqref{eig:mu}, we obtain the expression of $J_{\text{opt}}^{\mathcal C(k)}$ in \eqref{J:cycle}. To show $\bo W_{\text{opt}}^{\mathcal C(k)}\propto \bo A$, it suffices to show that
\begin{align}
\bo w_{\text{opt}}^{\mathcal C(k)}&\propto \bo \Omega^{-1} \bo G \bo \Gamma (\bo \Sigma+\bo \Gamma/P_\xi)^{-1}\bo h, \mbox{ } (\mbox{from  \eqref{theorem:ideal:weights}}) \nonumber \\
&\propto \bo \Omega^{-1} \bo G \bo 1_N, \mbox{ }  (\mbox{since } \bo h\propto \bo 1,\bo \Sigma\bo 1=\lambda\bo 1, \bo \Gamma\bo h=\mu\bo 1) \nonumber \\
&\propto \bo 1_L, \mbox{ }(\mbox{since } \bo G\bo 1_N\propto \bo 1_L,\bo \Omega^{-1}\bo 1_L\propto \bo 1_L) ,
\end{align}
which completes the proof.

\bibliographystyle{IEEEtran}
\bibliography{proposal/thesis}

\end{document}